\documentclass{jetpl}
\twocolumn

\lat

\title{Complex magnetic differentiation of cobalts in Na$_{x}$CoO$_{2}$ with 22~K N\'eel temperature}

\rtitle{Complex magnetic differentiation of cobalts in Na$_{x}$CoO$_{2}$\ldots}

\sodtitle{Complex magnetic differentiation of cobalts in Na$_{x}$CoO$_{2}$ with 22~K N\'eel temperature}

\author{I.\,R.\,Mukhamedshin$^{+*}$\/\thanks{e-mail: Irek.Mukhamedshin@kpfu.ru}, I.\,F.\,Gilmutdinov$^+$, M.\,A.\,Salosin$^+$, H.\,Alloul$^*$}

\rauthor{I.\,R.\,Mukhamedshin, I.\,F.\,Gilmutdinov, M.\,A.\,Salosin, H.\,Alloul}

\sodauthor{Mukhamedshin I.\,R., Gilmutdinov I.\,F., Salosin  M.\,A., Alloul H.}

\address{$^+$Institute of Physics, Kazan Federal University, 420008 Kazan, Russia\\~\\
$^*$Laboratoire de Physique des Solides, CNRS UMR 8502, Universit\'e Paris-Sud, 91405 Orsay, France, EU}

\dates{18 March 2014}{*}

\abstract{Single crystals of sodium cobaltates Na$_{x}$CoO$_{2}$ with $x \approx 0.8$ were grown by the floating zone technique. Using electrochemical Na de-intercalation method we reduced the sodium content in the as-grown crystals down to pure phase with 22~K N\'eel temperature and $x \approx 0.77$. The $^{59}$Co NMR study in the paramagnetic state of the $T_{N}=22$~K phase permitted us to evidence that at least 6 Co sites are differentiated. They could be separated by their magnetic behaviour into three types: a single site with cobalt close to non-magnetic Co$^{3+}$, two sites with the most magnetic cobalts in the system, and the remaining three sites displaying an intermediate behaviour. This unusual magnetic differentiation calls for more detailed NMR experiments on our well characterized samples.}

\PACS{71.27.+a, 76.60.-k}

\begin{document}

\maketitle

\textbf{Introduction.} - The influence of the dopant atoms on the electronic properties of conducting layers in complex layered oxides of transition elements is one of the most intriguing questions in strongly correlated electrons physics. This influence is under strong debates now in High Superconducting Temperature cuprates, while there are many experimental evidences in the sodium cobaltates Na$_{x}$CoO$_{2}$ for a large interplay between the Na atomic ordering and the electronic density on the Co sites. NMR/NQR is a powerful technique which allows to establish the relation between the local Na order and the local magnetic properties of the phases studied.

The rich phase diagram of sodium cobaltates~\cite{LangPRB2008} includes
ordered magnetic states~\cite{Mendels05}, high Curie-Weiss magnetism and
metal insulator transition~\cite{FooPRL}, superconductivity~\cite%
{TakadaNature} \emph{etc}. The Na$_{1}$CoO$_{2}$ compound is a band
insulator in which the Co sites are in filled-shell nonmagnetic Co$^{3+}$%
~states \cite{LangNa1}. At $x<1$, the system becomes metallic, but only
specific Na compositions can be obtained, which correspond to some Na
orderings as observed by diffraction techniques. For $x=0.5$ the Na atoms are ordered in an orthorhombic superstructure commensurate with the Co lattice. As a consequence small charge disproportionation into Co$^{3.5\pm \epsilon }$ with $\epsilon <0.2$ happens~\cite{Bobroff05}. In another peculiar phase with $x=2/3$ the ordering of the Na atoms on the hexagonal substructure is rather simple and results in a differentiation of the Co sites into non-magnetic Co$^{3+}$ sites and a metallic Kagome network of Co sites on which the doped holes are delocalized~\cite{EPL2009}.

Among the phases which ehibit a low $T$ magnetic order, a prevalent attention has been paid to the phase with a well defined N\'eel 
temperature of $T_{N}=22$~K which we had detected quite early by $\mu $SR~\cite{Mendels05}. This phase appears to be one of the most stable magnetic phases of sodium cobaltates, which has been found by many researchers. Most authors report that the Na content of this phase is close to the composition $x\approx 0.75$ while only one group insists on a composition of $x=0.825$~\cite{TaiwanPRB2013}.

Our recent $^{23}$Na NMR study of this sodium cobaltate phase with $T_{N}=22$~K ~\cite{Na077prb} allowed us to demonstrate that the two-dimensional structure of the Na order corresponds to 10 Na sites on top of a 13 Co sites unit cell, that is with $x=10/13\approx 0.77$. At the same time we have demonstrated the difficulties in producing homogeneous uniform samples of this cobaltate phase with $T_{N}=22$~K and have shown how $^{23}$Na NMR allows us to distinguish and separate the phase with $T_{N}=9$~K with higher sodium content. However, the stacking of the Na order and the associated Co charge disproportionation in this $T_{N}=22$~K phase is still unclear and requires a detailed $^{59}$Co NMR study on pure phase samples. This justifies the efforts undertaken here to synthesize new single crystal samples and to proceed a fine control of the Na content of the samples.

\textbf{Grystal growth.} - Single crystals of sodium cobaltates Na$_{x}$CoO$_{2}$ were grown by the floating zone technique in a new image furnace (Crystal Systems Corp., Japan) acquired in Kazan Federal University. At first stage polycrystalline samples of sodium cobaltates with the nominal composition Na$_{0.80}$CoO$_{2}$ were prepared using Na$_{2}$CO$_{3}$ (99,95\%, Alfa Aesar) and Co$_{3}$O$_{4}$ (99,7\%, Alfa Aesar) by the solid state reaction method. Carefully mixed powders were loaded into Al$_{2}$O$_{3}$ crucibles and sintered at 840~$^{\circ }$C for 12 hours. To obtain homogeneous materials the powders were twice reground and sintered at 860~$^{\circ }$C for 12 hours. To reduce the Na evaporation the `rapid heat-up' technique \cite{RapHeat} was used - the crucibles were inserted in the furnace preheated to the desired sintering temperature. To avoid absorption of moisture, intermediate dry grinding of the powder was done in Ar atmosphere. To form polycrystalline feed rods for the crystal growth process the Na$_{x}$CoO$_{2}$ powder was shaped into cylindrical bars of size 6--7~mm diameter and up to 150~mm length by pressing at an isostatic pressure of 25~MPa and then sintered at 1000~$^{\circ }$C for 12 hours in air.

\begin{figure}[tbp]
\centering
\includegraphics[width=1.0\linewidth]{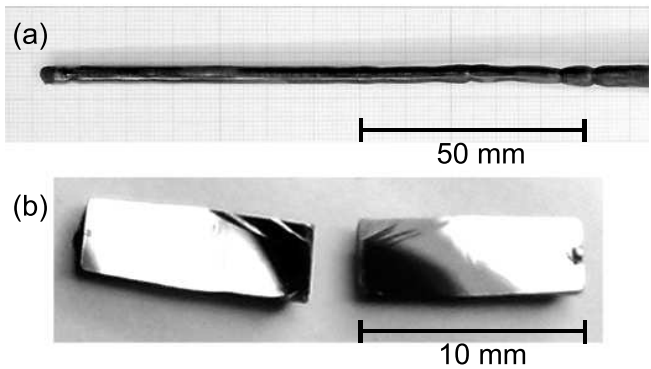}
\caption{Fig.~1. (a) Example of typical P2-phase sodium cobaltates crystal
ingot obtained by optical floating zone technique. (b) Two halves of an
as-grown crystal ingot freshly cleaved along the \textit{ab} planes surface.}
\label{FigCo77CentFitT2}
\end{figure}

After installing the sintered feed rod into the image furnace it was premelted with a mirror scanning velocity of 50 mm/h to densify it. After that a $\approx $ 20 mm long rod was cut and used as a first seed and hereafter the grown crystals were used as seeds. Many trials were done, in which crystals were grown in different atmospheric compositions and pressures. To obtain the P2-phase of sodium cobaltates we found that the best conditions were with four 300~W halogen lamps used as infrared radiation sources, growth rate of 7~mm/h with a pure oxygen flow of 50~ml/min and 4~atm pressure. The feed rod was rotated at 25 rpm and the growing crystal was not rotated. During this growth process the white powder of Na$_{2}$O is observed to deposit on the inner wall of the quartz tube making it less transparent, therefore the power applied to the lamps was gradually increased. With these conditions, during growth, we observed that the molten zone was rather stable and large Na$_{x}$CoO$_{2}$ single crystals with a sodium content $x\lesssim 0.8$ can be grown - Fig.~1(a) shows an example of such an as-grown typical crystal ingot. In such ingots the \textit{c}-axis is always perpendicular to the growth direction and the crystal is relatively easy to cleave along the \textit{ab} plane - see Fig.~1(b) as an example.

\textbf{Crystal characterization.} - Fig.~2(a) shows an example of the powder x-ray diffraction spectrum of the synthesized powder of sodium cobaltates with nominal composition Na$_{0.8}$CoO$_{2}$. Data collection for the diffraction patterns were performed on a Bruker D8 Advance powder diffractometer, using Cu K$\alpha $ radiation. Such spectra allow us to prove that the synthesized powder has a homogeneous Na content and do not exhibit any detectable Co$_{3}$O$_{4}$ or CoO impurity phases. Also the x-ray 00l diffraction pattern for the freshly cleaved as-grown crystal shown in Fig.~2(a) indicates that the Na$_{x}$CoO$_{2}$ crystal grown corresponds to a pure P2 phase with \textit{c}-axis lattice parameter  $c=$10.798(3)~\AA.

\begin{figure}[tbp]
\centering
\includegraphics[width=1.0\linewidth]{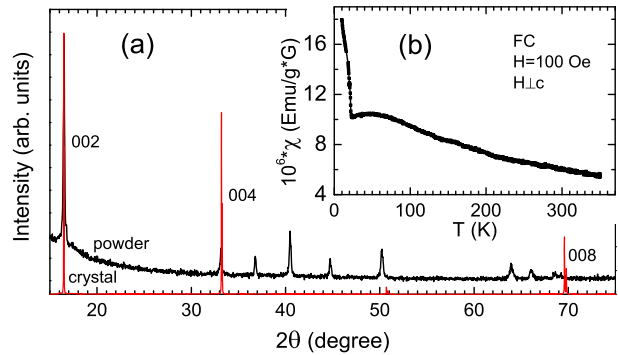}
\caption{Fig.~2. (Color online) (a) X-ray powder spectra for the sintered powder of Na$_{0.8}$CoO$_{2}$ (black line) and for the as-grown single crystal (red line). (b) Magnetization data taken with a VSM magnetometer on an as-grown single crystal sample. The data were taken after slow cooling (5~K/min) from room $T$ to 5~K in a field of 100~Oe and for $H \perp c$.}
\label{FigGrRawCryst}
\end{figure}

The temperature dependence of the bulk magnetization data taken with a VSM magnetometer (PPMS-9, Quantum Design) on an as-grown single crystal sample is shown in Fig.~2(b). It clearly exhibits a single magnetic transition at 22~K. But some change in the slope of magnetization T-dependence at about 9~K was also observed. Therefore, we carefully studied the 008 x-ray diffraction peak of the as-grown crystal - see Fig.~3(a). This diffraction pattern evidences two contributions, the dominant one from the known magnetic phase with $T_{N}=$22~K, and a small contribution of the $T_{N}=$9~K phase with slightly larger Na content, as we had shown in Ref.~\cite{Na077prb}. Also the $^{23}$Na NMR data shown in Fig.~3(b) clearly permits us to demonstrate the coexistence of the two phases in the as-grown crystal~\cite{Na077prb}.

We used the electrochemical Na de-intercalation method to reduce the sodium content in the crystals~\cite{NatureMat2010}. In the electrochemical cell the crystal sample was used as the working electrode, and counter and reference electrodes were Pt wires. A solution of 1M NaClO$_{4}$ in propylene carbonate was used as electrolyte. A home-built analogue potentiostat was used to control potentials and current in the cell. To reduce the sodium content in the crystals it is necessary to keep constant the negative potential difference between the reference and working electrodes for 10-20 hours. The complete disappearance of the $T_{N}=$9~K phase signals in x-ray diffraction spectra and in the $^{23}$Na NMR spectra was obtained with a -45~mV potential difference between the reference and working electrodes - see the corresponding spectra in Fig.~3(a) and (b).

Therefore, we reproducibly obtained high quality single crystals with pure
phase with $T_{N}=$22~K characterized well by x-ray diffraction bulk method and by $^{23}$Na NMR as a local probe. Moreover, the evaporation of Na during the crystal growth process, as well as the electrochemical sodium content reduction confirm that the sodium cobaltate phase with $T_{N}=$22~K has a sodium content smaller than 0.8, in agreement with our former determination of x=0.77~\cite{Na077prb}.

\begin{figure}[tbp]
\centering
\includegraphics[width=1.0\linewidth]{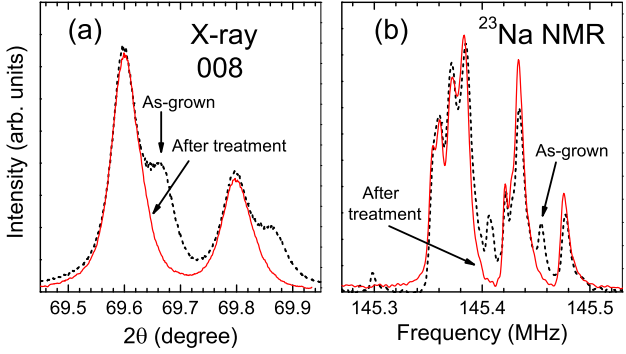}
\caption{Fig.~3. (Color online) (a) The 001 x-ray spectrum with (008) reflections for the as-grown crystal (black dashed line) which contains both $T_{N}=22$~K and $T_{N}=9$~K phases and pure $T_{N}=22$~K phase sample (red solid line) obtained after electrochemical treatment. (The double peak structure corresponds here to the two Bragg peaks associated with the Cu K$\protect\alpha _{1}$ and K$\protect\alpha _{2}$ radiations). (b) The $H\parallel c$ $^{23}$Na NMR spectrum, taken at T=80~K and B=12.897~T, of the as-grown crystal (black dashed line) and pure $T_{N}=22$~K phase sample (red solid line).}
\label{FigGrTreated}
\end{figure}

\textbf{$^{59}$Co NMR.} - The NMR measurements were carried out with a home-built coherent pulsed NMR spectrometer. The NMR spectra were taken \textquotedblleft point by point\textquotedblright\ with a $\pi /2-\tau -\pi$ radio frequency pulse sequence by varying the spectrometer frequency with equal frequency steps in a fixed external magnetic field. A Fourier mapping algorithm \cite{Clark} then has been used to construct the detailed NMR spectra. The minimal practical $\tau $ value used in our experiments was 5~$\mu s$.

In NMR the frequency of the central line which corresponds to the $-\frac{1}{2}\leftrightarrow \frac{1}{2}$ transition is determined by the applied field and the values of the magnetic shift $K$~\cite{H67_CoNMR}. Therefore, the different Co sites with different magnetic properties can be resolved in the NMR spectrum. In order to increase resolution a high magnetic field has been used. The Fig.~4(a) shows the central transitions part of the $H\parallel c$ $^{59}$Co NMR spectrum measured in the pure $T_{N}=$22~K single crystal in the high magnetic field $B_{0}$=12.82~T. As one can see this central line spectrum consists of \emph{at least} 7 lines. We have found that the line with the lowest frequency in Fig.~4(a) marked as (*) is not observable in all samples. It usually appears in the spectra of samples which were kept for some time in air. In the fresh prepared crystals this line is absent, therefore we believe that this NMR signal comes from some spurious phase. But all other peaks in Fig.~4(a) are present in all samples studied. Therefore, they certainly correspond to cobalt sites in the $T_{N}$=22~K phase of the sodium cobaltates.

\begin{figure}[tbp]
\centering
\includegraphics[width=1.0\linewidth]{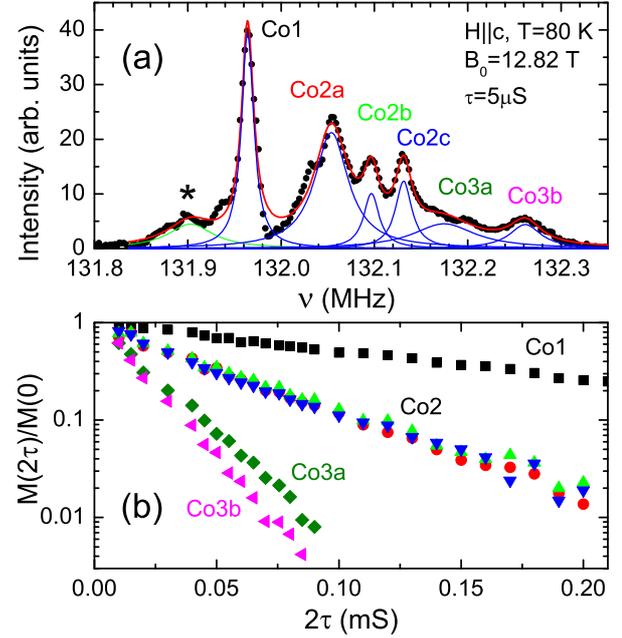}
\caption{Fig.~4. (Color online) (a) The spectrum of the $^{59}$Co NMR central lines measured at $\protect\tau $ = 5$\protect\mu $s and its decomposition into 7 Lorentz lines. (b) The transverse magnetization relaxation curves for each Co site.}
\label{FigCo77CentFitT2}
\end{figure}

We show also in Fig.~4(a) the result of the fitting procedure of the experimental spectrum by a sum of 7 Lorentzian lines. The parameters of the Lorentz function for the 6 lines which correspond to the cobalts in the $T_{N}$= 22~K sodium cobaltates phase are collected in Table~1. To make convenient comparison with the published data the positions of the lines were converted to the values of magnetic shifts $K_{ZZ}$ using the reference frequency value of 128.89~MHz (the gyromagnetic ratio is $\gamma /2\pi =$10.054~MHz/T for $^{59}$Co). Also the intensities (areas) of the 6 lines after $T_{2}$ correction were converted to relative intensities.

\begin{table}[tbp]
\caption{Table~1. The parameters of Co lines obtained after fitting the $H \parallel c$ $^{59}$Co NMR spectrum central line by a set of Lorentzian functions: $K_{ZZ}$ is the line center position converted to the magnetic shift value; $\Delta\protect\nu$ is the FWHM linewidth; $I$ is the relative intensity. The values of the transverse nuclear magnetization relaxation time $T_{2}$ for each cobalt site at $T$=80~K are also shown.}
\label{tab:Co77CoParams}
\begin{center}
\begin{tabular}{|c|c|c|c|c|}
\hline
Line & $K_{ZZ}$, \% & $\Delta\nu$, kHz & $I$, \% & $T_2$, $\mu $s \\ \hline
Co1 & 2.38(1) & 16(1) & 20(3) & 147(4) \\ 
Co2a & 2.45(1) & 44(2) & 32(3) & 45(2) \\ 
Co2b & 2.48(1) & 20(3) & 8(3) & 47(2) \\ 
Co2c & 2.51(1) & 21(3) & 14(3) & 44(3) \\ 
Co3a & 2.56(1) & 81(26) & 17(5) & 19(1) \\ 
Co3b & 2.62(1) & 41(10) & 9(3) & 16(1) \\ 
\hline
\end{tabular}
\end{center}
\end{table}

\textbf{Transverse magnetization relaxation.} We did determine the $^{59}$Co NMR central lines spectra similar to those of Fig.~4(a) with different values of the time interval $\tau $ between rf pulses. After that we integrated the spectra in the narrow frequency ranges around each peak position listed in the Table~1. The values of such integrals are decreasing with increasing $\tau $ values. That reflects the loss of the initial phase coherence of the nuclear spins and therefore the decrease of the transverse nuclear spin magnetization. In Fig.~4(b) the $M(t)/M(0)$ dependencies versus $2\tau $ for all 6 Co lines are shown. The decay of the transverse magnetization $M(t)$ as a function of time $t=2\tau $ was fitted by the equation $M(t)=M(0)exp(-t/T_{2})$, where $T_{2}$ is the transverse relaxation time. The $T_{2}$ values obtained for all central lines are listed in Table~1. It is clear that Co1 has the slowest relaxation whereas the Co3a and Co3b are fast relaxing signals. The NMR data for the Co2a, Co2b and Co2c sites have nearly the same $T_{2}$ relaxation time. Therefore, the specific phase of sodium cobaltates Na$_{x}$CoO$_{2}$ with 22~K N\'{e}el temperature is characterized by three types of cobalts: Co1 is close to the non-magnetic Co$^{3+}$ charge state, Co3a and Co3b are the most magnetic cobalts in the system, and the Co2a, Co2b, Co2c sites display an intermediate behaviour.

\textbf{Discussion}.- In one of the earliest papers of $^{59}$Co NMR study of single crystals Na$_{0.75}$CoO$_{2}$ with $T_{N}=22$~K the presence of 3 distinct cobalt signals Co1, Co2 and Co3 was reported~\cite{MHJ075prl}. There it was clear that the Co1 signal was quite close to that expected for a Co$^{3+}$ charge state and involved about 30(4)\% of the total number of sites. Two other signals, Co2 and Co3, correspond to "magnetic" cobalts. There, as we did in our initial work on what later appeared to be the $x=2/3$ phase \cite{CoPaper}, the authors did their analysis of the $^{59}$ Co NMR spectra, using the quadrupolar satellites for $H||c$ and for the so-called \textquotedblright magic angle\textquotedblright orientation of the applied field. Furthermore, the
single phase content was not established for their samples. Therefore, the experimental conditions were not clean enough to resolve the 6 sites we detected in the present work.

The $^{23}$Na NMR central line study reported in the recent paper \cite{TaiwanPRB2013} completely reproduces the results of the previous NMR studies of the sodium cobaltate phase with $T_{N}=22$~K~\cite{Na077prb,MHJ075prl}. Also the $^{59}$Co NMR central line spectrum reported in that paper is quite similar to our spectrum presented in Fig.~4(a) except that the authors of Ref.~\cite{TaiwanPRB2013} have missed the two fast relaxing signals Co3a and Co3b that we have studied and identified in all our samples. Though Ref.~\cite{TaiwanPRB2013} does not contain the necessary experimental details, we can envision that the minimal interval $\tau $ between rf pulses in their experiments was too long, so that the fast relaxing NMR signals died during their spectrometer so called "dead time". As a consequence the structure proposed in Ref.~\cite{TaiwanPRB2013} for the $T_{N}=22$~K phase based on NMR spectra in which two major Co sites are missing is totally unreliable.

\textbf{Conclusion}.- This NMR work has permitted us to evidence that at least 6 Co sites are differentiated in the paramagnetic state of this $x=0.77$ phase. We have shown that the actual magnetic properties apparently differentiate only three distinct behaviours. We recall that for $x=2/3$ NMR shows four inequivalent Co sites which demonstrate only two types of magnetic behaviour. In that case we could complete a full determination of the atomic structure and could associate the two distinct magnetic sites with the two sites of the Kagom\'{e} structure of the Co plane~\cite{EPL2009}. Obviously the magnetic difference was associated then with a charge disproportionation, the Co1 sites being weakly magnetic with a charge near Co$^{3+}$, while the holes were delocalized mostly between the Co2 sites.

In the present situation, for  $x \approx 0.77$, we do not have at this stage enough information to conclude whether the three magnetic behaviours detected correspond to different charge disproportionations on three types of sites. By analogy with $x=2/3$, the Co1 sites could be nearly Co$^{3+}$ sites and holes should delocalize on the Co3 sites which exhibit the most magnetic behaviour. The detected  Co2 sites could be associated with an independent charge disproportionation on the Co2 sites but could result as well from transferred hyperfine couplings with the Co3 sites in the lattice structure. To better understand the electronic properties of these three types of Co sites, independent determinations of the charge disproportionation are necessary. 

The fact that we have now established sample synthesis methods which allow us to control the Na content and order in this phase will permit us to perform more detailed NMR experiments to characterize fully both the charge disproportionation and the magnetic properties and to correlate them with each other. Such a work program is presently underway and might help us altogether to understand why the ground state properties of these two phases are so different.

We gratefully thank A.~N.~Lavrov for consulting on the crystal growth technique, F.~Bert, and P.~Mendels for their help on the experimental NMR techniques, and A.~V.~Dooglav for helpful discussions and help in manuscript preparation. I.R.M. thanks the European Science Foundation for the grant (INTELBIOMAT - 4211) which permitted short exchange visits in Orsay. This study was partly supported by the RFBR under project 14-02-01213a. I.F.G. thanks for support the Ministry of Education and Science of the Russian Federation.

\end{document}